\documentclass[10pt,twocolumn]{IEEEtran}
\usepackage{epsfig,latexsym}
\usepackage{float}
 \usepackage{enumerate}
\usepackage{indentfirst}
\usepackage{amsmath}
\usepackage{amssymb}
\usepackage{times}
\usepackage{subfigure}
\usepackage{cite}
\usepackage{url}
\usepackage{color}
\usepackage{setspace}

\definecolor{green}{rgb}{0.796,0.948,0.816}

\begin{document}
\title{Robust Beamforming Techniques for Non-Orthogonal Multiple Access Systems with Bounded Channel Uncertainties}
\author{Faezeh Alavi, Kanapathippillai Cumanan, Zhiguo Ding and Alister G. Burr
}
\maketitle

\begin{abstract}
In this letter, we propose a robust \mbox{beamforming} \mbox{design}
for non-orthogonal multiple access (NOMA) based multiple-input single-output (MISO)
downlink systems.
In \mbox{particular}, the robust power minimization problem is studied with imperfect channel state information (CSI),
where the beamformers are designed by incorporating norm-bounded channel uncertainties to provide the required quality
of service at each user. This robust scheme is developed based on the worst-case performance optimization framework.
In terms of beamforming vectors, the original robust design is not convex and therefore, the robust beamformers cannot
be obtained directly. To circumvent this non-convex issue, the original intractable problem is reformulated into a
convex problem, where the non-convex constraint is converted into a linear matrix inequality (LMI) by exploiting S-Procedure.
Finally, simulation results are
provided to demonstrate the effectiveness of the proposed robust
design.
\end{abstract}


\vspace{-0.5cm}
\section{Introduction}
%
Non-orthogonal multiple access (NOMA) is a promising multiple access technique for 5G networks
which has the potential to address the 
issues associated with the exponential growth of data traffic
such as spectrum scarcity and massive connectivity \cite{6730679,6868214,7273963,7676258,7263349}.
In contrast to conventional multiple access schemes,
NOMA allows different users to efficiently share the same resources (i.e., time, frequency and code)
at different power levels so that the user with lower channel gain is served with a higher power and vice versa.
In this technique, a successive interference cancellation (SIC) approach is employed at receivers to separate multi-user
signals, which significantly enhances
the overall spectral efficiency.
In other words,
NOMA has the capability to control the
interference by sharing resources
while increasing system throughput
with a reasonable additional complexity \cite{7263349}.

Recently, a significant amount of research has focused in studying
several practical issues in NOMA scheme.
In particular, beamforming designs for multiple antenna NOMA networks have received  
a great deal of interest in the research community due to their additional degrees of freedom and diversity gains \cite{7433470,7095538,7277111}. 
A general framework for a multiple-input multiple-output (MIMO) NOMA system has been developed for both the downlink
and the uplink in \cite{7433470} whereas 
the throughput maximization problem was studied for a two-user MIMO NOMA system in \cite{7095538}.
The sum  rate maximization problem for a multiple-input single-output (MISO) NOMA has been investigated in \cite{7277111}
through the minorization maximization algorithm. 
In most of the existing work, beamforming designs have been proposed  
for NOMA schemes with the assumption of perfect channel state information (CSI) at the transmitter \cite{7433470,7095538,7277111}.
However, 
this assumption might not be always valid for practical scenarios 
due to channel estimation and quantization errors \cite{4700145,4895658,5073854,5074429,5665898,6967767,7047328}. On the other hand, channel uncertainties significantly influence the performance of the
SIC based receivers as the 
decoding order of the received multi-user signals is determined with respect to the users' effective channel gains.
Therefore, it is important to take into account the channel
uncertainties especially in the beamforming design for NOMA networks.
Motivated by this practical constraint, we
focus on robust beamforming design based on the worst-case performance optimization framework
to tackle the norm-bounded channel uncertainties \cite{5755208,6156468,7177089,7547398}.
In \cite{5755208}, the robust beamforming design has been developed
for providing secure communication in wireless networks with   
imperfect CSI.
By incorporating the 
bounded channel uncertainties, the robust sum power minimization
problem 
is investigated in \cite{6156468}
for a downlink multicell network with the worst-case signal-to-interference-plus-noise-ratio (SINR) constraints
whereas the robust weighted sum-rate maximization was studied for multicell downlink MISO systems in \cite{7177089}.
In \cite{7547398}, a robust minimum mean square error based beamforming technique is
proposed for multi-antenna relay channels with imperfect CSI between the relay and the users.
In the literature, there are two types of NOMA schemes considered:
I) clustering NOMA \cite{6735800,7015589,7442902}, II) non-clustering NOMA \cite{5755208,6156468,7177089,7547398}.
In the clustering NOMA scheme, all the users in a cell are grouped into $N$ clusters with two users
in each cluster, for which a transmit beamforming vector is designed to support those two users
through conventional multiuser beamforming designs. The users in each cluster are supported by a NOMA
beamforming scheme. However, in the non-clustering NOMA scheme, there is no clustering and each
user is supported by its own NOMA based beamforming vector.
In \cite{7442902}, the authors studied a robust NOMA scheme for the MISO channel
to maximize the worst-case achievable sum rate with a
total transmit power constraint.

In this letter, we follow the second class of research where NOMA scheme applied between all users and there is the spectrum
sharing between all users in cell. Then, we propose a robust beamforming design for NOMA-based MISO
downlink systems.
In particular, the robust power minimization problem is solved
based on worst-case optimization framework to provide the required quality of service at each user
regardless of the associated channel uncertainties.
By exploiting \mbox{S-Procedure,} the original \mbox{non-convex} problem is
converted into a convex one by recasting the non-convex constraints into linear matrix inequality (LMI) forms.
Simulation results are provided
to validate the \mbox{effectiveness} 
of the robust design by comparing the performance of the robust scheme with that of the non-robust approach.
{The work in \cite{7442902} also studied the worst-case based robust scheme
for MISO NOMA system, however, there are main differences between our proposed scheme and the work in \cite{7442902}.
A clustering NOMA scheme is developed in \cite{7442902} by grouping
users in each cluster.
In this scheme, a single beamformer is designed to transmit the signals for all users
in the same cluster whereas, in this letter, the signal for each user is transmitted with a dedicated beamformer.
In addition, both beamforming designs are completely different as the work in \cite{7442902}
proposes robust sum-rate maximization based design whereas this letter solves robust power minimization problem
with rate constraint on each user. In terms of solutions, the work in \cite{7442902}
exploits the relationship between MSE and achievable rate and derives an equivalent non-convex problem, which
is decoupled into four sub-problems and those problems are iteratively solved to realize the solution of the original problem.
In this letter, the robust power minimization problem is formulated by deriving the worst-case achievable rate.
The original problem formulation turns out to be non-convex and we exploit S-Procedure and semidefinite
relaxation to convert it to a convex one. Hence, the work in \cite{7442902} and the proposed work in
this letter are different including problem formulation and the solution approaches.}

\vspace{-0.2cm}
\section{System Model and Problem Formulation}
We consider NOMA-based downlink transmission where a
base station (BS) sends information to $K$ users 
$U_1,U_2, \ldots, U_K$.
It is assumed that the BS is equipped with $M$ antennas whereas each user consists of a single antenna. 
The channel coefficient vector between the BS and the $k^{th}$ user $U_k$ is denoted by $\mathbf {h}_k \in \mathbb{C}^{M \times 1}~ (k = 1,\ldots,K)$
and $\mathbf{w}_{k}\in \mathbb{C}^{M \times 1}$  represents the corresponding beamforming vector of the $k^{th}$ user $U_{k}$.
The received signal at $U_k$ is given by
\begin{equation}\label{signal}
  y_k = \mathbf {h}_k^H \mathbf{w}_k s_k + \sum_{m\neq k}\mathbf{h}_k^H \mathbf {w}_m  s_m + n_k, \qquad \forall k,
\end{equation}

\vspace{-0.3cm}
\noindent where $s_{k}$ denotes the symbol intended for $U_{k}$
and \mbox{$n_k \sim \mathcal{CN }(0, \sigma_k^2)$} represents a zero-mean additive white Gaussian noise with variance $\sigma_k^{2}$.
The power of the symbol $s_{k}$ is assumed to be unity, i.e., $\mathbb{E}(|s_k|^2)=1$. In practical scenarios, it is difficult to provide perfect CSI at the transmitter due to  
channel estimation and quantization errors. Therefore, we consider a robust beamforming design to overcome these channel uncertainties.
In particular, we incorporate norm-bounded channel uncertainties in the design as

\vspace{-0.5cm}
\begin{align} \label{delta}
& \mathbf{{h}}_k=\mathbf{\hat{h}}_k +\Delta\mathbf{\hat{h}}_k,\quad \|\Delta\mathbf{\hat{h}}_k\|_2=\|\mathbf{{h}}_k-\mathbf{\hat{h}}_k\|_2 \leq \epsilon, 
\end{align}

\vspace{-0.25cm}
\noindent where $\mathbf{\hat{h}}_k$, $\Delta\mathbf{\hat{h}}_k$ and
$\epsilon \geq 0$ denote the estimate of $\mathbf{{h}}_k$, the norm-bounded channel estimation error and the channel estimation error bound, respectively.

In the NOMA scheme, user multiplexing is performed in the power domain
and the SIC approach is employed at receivers to separate signals between different users.
In this scheme, users are sorted
based on the norm of their channels,
i.e., $\|\mathbf{h}_{1}\|_2\leq\|\mathbf{h}_{2}\|_2\leq\ldots\leq\|\mathbf{h}_{K}\|_2$.
For example, the $k^{th}$ user decodes the signals intended for the users from $U_{1}$ to $U_{k-1}$
using the SIC approach 
whereas the signals intended for the rest of the users (i.e., $U_{k+1},\ldots,U_{K}$) are treated as interference at the $k^{th}$ user.
Based on this SIC approach, the $l^{th}$ user can detect and remove
the $k^{th}$ user's signals for $ 1 \leq k < l $ \cite{7277111}.
Hence, the signal at the $l^{th}$ user after removing the first $k-1$ users' signals to detect the $k^{th}$ user is represented as

\vspace{-0.6cm}
\begin{align}\nonumber
y_l^k=\mathbf {h}_l^H \mathbf{w}_k s_k+\sum_{m=1}^{k-1} \Delta\mathbf{\hat{h}}_l \mathbf{w}_m s_m+\sum_{m=k+1}^{K}\mathbf{h}_l^H \mathbf {w}_m s_m+n_l,&&
\end{align}

\vspace{-0.7cm}
\begin{eqnarray} \label{signal2}
 \qquad\qquad \qquad\qquad\quad\quad\forall k, \, l\in\{k,k+1,\ldots,K\},&&
\end{eqnarray}

\vspace{-0.3cm}
\noindent
where the first term is the desired signal to detect $s_k$ and 
the second term is due to imperfect CSI at the receivers during 
the SIC process. 
Due to the channel uncertainties,  the signals intended for the users $U_{1},\ldots,U_{k-1}$ cannot be completely removed by the $l^{th}$ user. The third term is the interference
introduced by the signals intended to the users $U_{k+1},\ldots,U_{K}$. 
According to the SIC based NOMA scheme, the $l^{th}$ user should be able to detect
all $k^{th}\:(k<l)$ user signals. Thus,
the achievable rate of $U_k$ can be defined as follows:

\vspace{-0.4cm}
\begin{align}\label{SINR}
R_k=\log_2 \Big(1+\min_{l \in \{k, k+1, \ldots, K\}} ~ \text{SINR}^k_l\Big),
\end{align}

\vspace{-0.2cm}
\noindent where $\text{SINR}^k_l$ 
denotes the SINR 
of the $k^{th}$ user's signal at the $l^{th}$ user which can be written as
\begin{align}\label{SINR2}
&\text{SINR}_l^k= \dfrac{\mathbf{h}_l^H \mathbf{w}_{k} \mathbf{w}_{k}^H \mathbf{h}_l}{\displaystyle{ \sum_{m=1}^{k-1}}\Delta\mathbf{\hat{h}}_l^H \mathbf{w}_{m}\mathbf{w}_{m}^H \Delta\mathbf{\hat{h}}_l+\displaystyle{\sum_{m=k+1}^{K}}\mathbf{h}_l^H \mathbf{w}_{m}\mathbf{w}_{m}^H\mathbf{h}_l+\sigma_l^2}.
\end{align}

\vspace{-0.3cm}
For this network setup, we study robust power minimization by incorporating channel uncertainties to satisfy the required SINR at each user. This robust beamforming design is developed by considering the worst-case SINR of each user, which can be formulated as

\vspace{-0.6cm}
\begin{subequations}\label{mainproblem}
\begin{align}
&\min_{\mathbf{w}_k\in \mathbb{C}^{M\times1}} \sum_{k=1}^{K} \| \mathbf{w}_k\|^2_2, \\ \label{const1.1}
&\mathbf{s.t.} \min_{\|\Delta\mathbf{\hat{h}}_l\|_2 \leq \epsilon} \hspace{-0.1cm}\big(\min_{l \in \{k, k+1, \ldots, K\}} \hspace{-0.12cm} \text{SINR}^k_l\big)\geq \gamma_k^{min},\;\,  \forall k,
\end{align}
\end{subequations}

\vspace{-0.2cm}
\noindent where $\gamma_k^{min}=(2^{R_k^{min}}-1)$ is the minimum required SINR to achieve a target rate $R_k^{min}$ at $U_k$.

\vspace{-0.3cm}
\section{Robust Beamforming Design}
The problem formulation in \eqref{mainproblem} is not convex and the optimal robust beamformers cannot be obtained directly. To tackle this issue, we introduce a new matrix variable $\mathbf{W}_k=\mathbf{w}_k \mathbf{w}_k^H$
and reformulate the original robust problem in \eqref{mainproblem} into the following optimization framework without loss of generality:

\vspace{-0.5cm}
\begin{subequations}\label{Procedure}
\begin{align}
&\min_{\mathbf{W}_k\in \mathbb{C}^{M\times M}}  \sum_{k=1}^{K} {\rm Tr}(\mathbf{W}_k), \\ \label{Procedure2.1}
& \mathbf{s.t.} \;\; \varpi_{kl},~ \quad \forall k, ~l=k,\ldots,K,\\ \label{Procedure2.3}
&\qquad\mathbf{W}_k \succeq \mathbf{0},~ \text{rank}(\mathbf{W}_k)=1, \qquad \forall k,
\end{align}
\end{subequations}
where $\varpi_{kl}$ is defined in Appendix \ref{App.A}.

However, the reformulated problem in \eqref{Procedure} is still not convex
for two reasons; the rank-one constraint and unknown channel uncertainties, i.e., $\Delta\mathbf{\hat{h}}_k$,
which lead to an intractable problem.
The rank-one constraint in \eqref{Procedure2.3}  
can be relaxed by exploiting semi-definite relaxation (SDR).
To remove the unknown channel uncertainties and solve the original problem with available knowledge of imperfect CSI (error bound), we employ \mbox{S-procedure} to recast the non-convex constraints into LMIs. 

\textit{Lemma 1:}
By relaxing the rank-one constraints on $\mathbf{W}_k$, the original problem in \eqref{Procedure} can be recast into
the following convex problem: 

\vspace{-0.8cm}
\begin{subequations}\label{Procedure2}
\begin{align}
& \min_{ \tiny {\begin{array}{l l} \mathbf{W}_k\in \mathbb{C}^{M\times M},\\
 \lambda_{kl}\geq 0 \end{array}} } \sum_{k=1}^{K} {\rm Tr}(\mathbf{W}_k), \\ \label{Procedure22.1}
& \mathbf{s.t.}~ 
\mathbf{W}_k \succeq \mathbf{0},~\mathbf{C}_{kl}\succeq \mathbf{0}, \quad \forall k, ~l=k,\ldots,K 
\end{align}
\end{subequations}

\vspace{-0.2cm}
\noindent where $\mathbf{C}_{kl}$ is defined in Appendix \ref{App.B}.

\begin{IEEEproof}
 Please refer to Appendix \ref{App.B}.
\end{IEEEproof}

The problem in \eqref{Procedure2} 
is a standard semidefinite \mbox{programming} (SDP) and can be efficiently solved using interior-point \mbox{methods.}
The optimal solution for the original problem in \eqref{mainproblem} can be
obtained through extracting the eigenvector corresponding to the maximum
eigenvalue of the rank-one solution of \eqref{Procedure2}.
Thus, the following lemma
holds to show that the optimal solution to \eqref{Procedure2} is rank one.

\textit{Lemma 2:}
Provided the problem in \eqref{Procedure2} is feasible, there always exists a rank-one optimal solution $\{\mathbf{W}^*_k\}$.
%
%

\begin{IEEEproof}
 Please refer to Appendix \ref{App.C}.
\end{IEEEproof}

\begin{figure}[t!]
  \begin{center}
    \includegraphics[scale=0.32]{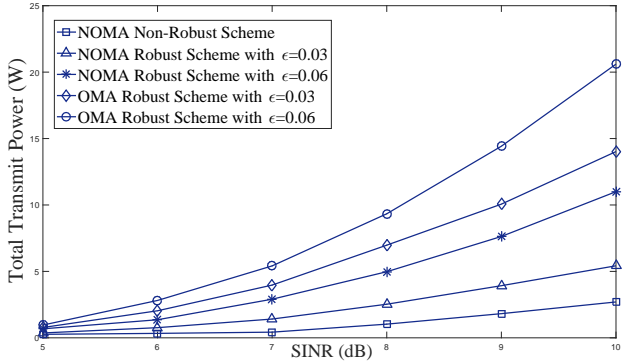} 
    \vspace{-0.3cm}
    \caption{Total transmit power versus different SINR thresholds for the robust and non-robust schemes with different channel estimation \mbox{error bounds, $\epsilon$.} }\vspace{-0.7cm}
   \label{pr2}
    \end{center}
\end{figure}

\section{Simulation Results}
To assess the performance of the proposed robust beamforming approach, we consider a single cell downlink transmission, where a multi-antenna BS serves {single-antenna users which are uniformly distributed over the circle with a radius of 1000 meters around the BS, but no closer than $d_0=100$ meters. }
{The small-scale fading of the channels is Rayleigh which represents an isotropic scattering environment. We model the large-scale fading effects as the product of path loss and shadowing fading. The log-normal shadowing is considered with standard deviation $\sigma_{0}=8$ dB, scaled by $(\frac{d_k}{d_0})^{-\beta}$ to incorporate the path-loss effects where $d_{k}$ is the distance between $U_k$ and the BS, measured in meters and $\beta=3.8$ is the path-loss exponent. Throughout the simulations, it is assumed that the BS is equipped with eight antennas ($M = 8$) and it serves three users ($K = 3$). The noise variance at each user is assumed to be $0.01$ (i.e., $\sigma_{k}^{2} = 0.01$)  
and the target rates for all users are the same. }
The term ``Non-robust scheme" refers to the scheme where
{the BS has imperfect CSI without any information on the channel uncertainties and the beamforming
vectors are designed based on imperfect CSI without incorporating channel uncertainty information.}

\begin{figure}[t!]
  \begin{center}
    \includegraphics[height=2.1in,width=3.6in]{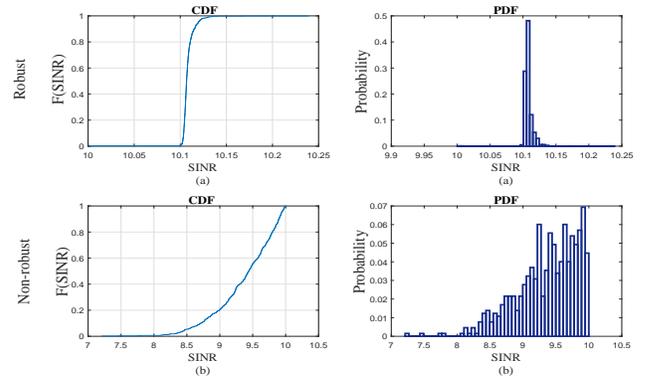} 
    \vspace{-0.7cm}
    \caption{Comparison CDF and PDF of minimum achieved SINR for (a) the \mbox{robust} scheme and (b) the non-robust scheme with $\epsilon=0.06$, \mbox{$\gamma_k^{min}=10$ dB.}}\vspace{-0.2cm}
   \label{cdf}
   \vspace{-0.6cm}
  \end{center}
\end{figure}

First, we study the impact of channel uncertainties on the required total transmit power.
Fig. \ref{pr2} depicts the required total transmit power against
different SINR thresholds for the robust and the non-robust NOMA schemes as well as OMA scheme with different error bounds.
As seen in Fig. \ref{pr2},
the robust scheme requires more transmit power
than that of the non-robust scheme.
This is because the robust scheme satisfies the required
SINR all the time, at the price of more transmit power at the BS
whereas the non-robust scheme does not.
The difference between the required transmit power for the robust and the non-robust schemes increases with error bounds.
This is because incorporating all possible sets of errors in the beamforming design to satisfy high SINR thresholds
requires more transmit power in the robust scheme.
Moreover, as seen in \mbox{Fig. \ref{pr2}}, the conventional framework, orthogonal multiple access (OMA), requires
more transmit power to achieve the same rate in comparison with NOMA scheme. This demonstrates that the NOMA scheme
yields a better performance in terms of spectral and energy efficiencies.

Next, we evaluate the performance of the proposed robust and non-robust schemes in terms of the minimum achieved SINR between users.
Fig.  \ref{cdf} provides cumulative \mbox{distribution} function (CDF) and probability density function (PDF) \mbox{obtained} from $1000$ random sets of channels with error bounds of $0.06$  $(\epsilon = 0.06)$
where the SINR threshold has
been set to $10$ dB at each user.
As evidenced by the results, the robust scheme outperforms the non-robust scheme
in terms of minimum achieved SINRs. In addition, the robust scheme satisfies the SINR thresholds all the time
regardless of the channel uncertainties whereas
 the non-robust design fails to
satisfy the minimum SINR requirements. 
%

\vspace{-0.5cm}
\section{Conclusion}

\vspace{-0.1cm}
In this letter, we propose a robust beamforming
design for the downlink of a NOMA based MISO network
by taking into account the norm-bounded channel uncertainties.
However, the original robust problem formulation is not convex due to the imperfect CSI.
To cope with this challenge, we exploited \mbox{S-procedure}
to reformulate the original non-convex problem into a convex optimization
framework by recasting the original non-convex constraints into an LMI form.
Simulation results demonstrate that the
proposed robust scheme offers a better performance than the non-robust approach
by satisfying the SINR requirement at each user all the time regardless of associated channel uncertainties.

\appendices \numberwithin{equation}{section}
\setcounter{equation}{0}
\vspace{-0.2cm}
\section{Derivation of $\varpi_{kl}$} \label{App.A}
%
\vspace{-0.15cm}
The equivalent transformations of \eqref{const1.1} can
be obtained as $\varpi_{kl}$ as follows:

\vspace{-0.5cm}
\begin{align} \nonumber
&\!\!\!\!\left\{\begin{array}{l l} \!\!\!\!\displaystyle{\min_{\|\Delta\mathbf{\hat{h}}_k\|_2 \leq \epsilon}}
\!\Big(\frac{\mathbf{h}^H_k \mathbf{W}_k \mathbf{h}_k}{\displaystyle{\!\sum_{m=1}^{k-1}}\!\!\Delta\mathbf{\hat{h}}_k^H \mathbf{W}_m \Delta\mathbf{\hat{h}}_k \!+\!\!\!\sum_{m=k+1}^{K}{h}^H_k \!\!\!\mathbf{W}_m \mathbf{h}_k +\sigma_k^2}\Big)
\!\!\leq \!\!\gamma^{min}_k\!\!\!, \\
\!\!\!\!\!\displaystyle{\min_{\|\Delta\mathbf{\hat{h}}_{k+1}\|_2 \leq \epsilon}}
\!\!\Big(\frac{\mathbf{h}^H_{k+1}\mathbf{W}_k\mathbf{h}_{k+1}}{\displaystyle{\!\!\sum_{m=1}^{k-1}}\!\!\Delta\mathbf{\hat{h}}_{k+1} \mathbf{W}_m \Delta\mathbf{\hat{h}}_{k+1}\!\!+\!\!\hspace{-0.3cm}\sum_{m=k+1}^{K}\!\!\!\!\!\mathbf{h}^H_{k+1} \mathbf{W}_m \mathbf{h}_{k+1}\!\!+\!\!\sigma_{k+1}^2}\!\Big)\\
\hspace{7.5cm}\leq \!\!\gamma^{min}_k\!\!\!, \\
\vdots\\
\!\!\!\!\displaystyle{\min_{\|\Delta\mathbf{\hat{h}}_K\|_2 \leq \epsilon}}
\!\Big(\frac{\mathbf{h}^H_K \mathbf{W}_k \mathbf{h}_K}{\displaystyle{\!\!\sum_{m=1}^{k-1}}\Delta\mathbf{\hat{h}}_K \mathbf{W}_m \Delta\mathbf{\hat{h}}_K +\!\!\!\!\!\sum_{m=k+1}^{K}\!\!\!\!\mathbf{h}^H_K \mathbf{W}_m \mathbf{h}_K \!\!+\!\!\sigma_K^2}\!\Big)
\!\!\leq \!\!\gamma^{min}_k\!\!\!,  \end{array}
\right.
\\ \nonumber
&\Leftrightarrow \!\!\!\displaystyle{\min_{\|\Delta\mathbf{\hat{h}}_l\|_2 \leq \epsilon}}
\Big(\frac{\mathbf{h}^H_l \mathbf{W}_k \mathbf{h}_l}{\displaystyle{\sum_{m=1}^{k-1}}\Delta\mathbf{\hat{h}}_l^H \mathbf{W}_m \Delta\mathbf{\hat{h}}_l +\!\!\!\hspace{-0.2cm}\sum_{m=k+1}^{K}\!\!\!\!\mathbf{h}^H_l \mathbf{W}_m \mathbf{h}_l\!+\!\sigma_l^2}\!\Big)
\!\!\leq \!\!\gamma^{min}_k \\ \label{varpi}
&\hspace{1.5cm} \triangleq \varpi_{kl},\;\; \forall k,~ l=k,\ldots,K.
\end{align}

\vspace{-0.8cm}
\section{Proof of Lemma 1 } \label{App.B}

\vspace{-0.25cm}
To incorporate the channel uncertainties in the robust
optimization framework, we exploit S-procedure 
to convert the non-convex constraint into LMI form.
%
%
By applying \mbox{S-procedure} \cite{sproceturebook}, the constraint \eqref{Procedure2.1} is \mbox{derived as}
\begin{align}\nonumber
& \Delta\mathbf{\hat{h}}_l^H \mathbf{I} \Delta\mathbf{\hat{h}}_l -\epsilon^2 \leq 0 \Rightarrow \Delta\mathbf{\hat{h}}_l^H (\sum_{m\neq k} \mathbf{W}_m - \mathbf{W}_k/\gamma^{min}_k )\Delta\mathbf{\hat{h}}_l
\end{align}

\vspace{-0.5cm}
\begin{align} \nonumber
&+2{\rm{Re}}\{\mathbf{\hat{h}}_l^H(\sum_{m=k+1}^{K} \hspace{-0.2cm}\mathbf{W}_m -\mathbf{W}_k/ \gamma^{min}_k)\Delta\mathbf{\hat{h}}_l \}+\mathbf{\hat{h}}_l^H
(\hspace{-0.2cm}\sum_{m=k+1}^{K}\hspace{-0.2cm}\mathbf{W}_m\\
& - \mathbf{W}_k/ \gamma^{min}_k )\mathbf{\hat{h}}_l + \sigma_l^2 \leq 0,
\end{align}

\vspace{-0.1cm}
Then, the constraint \eqref{Procedure2.1} can be reformulated with \mbox{$\lambda_{kl}\geq 0$} as the following semidefinite constraint
\begin{align}\label{const1}
&\!\!\!\!\mathbf{C}_{kl}= \left [\begin{array}{l l}
\hspace{-0.1cm}\lambda_{kl} \mathbf{I}+\mathbf{\phi}_k+\mathbf{\nu}_k & \!\!\!\mathbf{\phi}_k \mathbf{\hat{h}}_l \\
\hspace{-0.1cm}\mathbf{\hat{h}}_l^H \mathbf{\phi}_k & \!\!\!\mathbf{\hat{h}}_l^H \mathbf{\phi}_k \mathbf{\hat{h}}_l-\sigma_k^2-\lambda_{kl}\epsilon^2\!\!\!
\end{array}
\right] \succeq \mathbf{0},
\end{align}
where \mbox{$\mathbf{\phi}_k=\frac{\mathbf{W}_k}{\gamma^{min}_k}-\sum_{m=k+1}^{K}\mathbf{W}_m$ and $\mathbf{\nu}_k=-\sum_{m=1}^{k-1}\mathbf{W}_m$.}

This completes the proof of Lemma 1. \qquad\qquad\qquad $\blacksquare$

\section{Proof of Lemma 2 } \label{App.C}

\vspace{-0.2cm}
To prove Lemma 2, we examine the Karush-Kuhn-Tucker
(KKT) conditions of \eqref{Procedure2}. First, let 
$\mathbf{Y}_{k}\in \mathbb{C}^{M\times M}$, \mbox{$\mathbf{T}_{kl}\in \mathbb{C}^{(M+1)\times(M+1)}$} and $\mu_{kl}\in \mathbb{R}_{+}$
denote the dual variable of the
constraints in \eqref{Procedure22.1}, 
respectively. Then, the Lagrangian
dual function of \eqref{Procedure2} can be written as

\vspace{-0.6cm}
\begin{align}\nonumber
&\mathcal{L}(\mathbf{W}_k,\lambda_{kl},\mathbf{T}_{kl},\mu_{kl},\mathbf{Y}_k)\!=\!\sum_k {\rm Tr}(\mathbf{W}_k)\!-\!\sum_k {\rm Tr}(\mathbf{Y}_k\mathbf{W}_k)-\\ \label{1}
&\!\!\!\!\!\sum_{k,l}\!{\rm Tr}(\mathbf{T}_{kl} \mathbf{A_1})\!-\!\!\sum_{k,l}\!{\rm Tr}[\mathbf{T}_{kl} \mathbf{H}^H_{l} \mathbf{\phi}_k \mathbf{H}_{l}]\!\!-\!\!\sum_{k,l}\!{\rm Tr}(\mathbf{T}_{kl} \mathbf{A_{2}}),
\end{align}

\vspace{-0.25cm}
\noindent where $\mathbf{H}_{l}=[\mathbf{I} \quad \mathbf{h}_l]$ and

\vspace{-0.5cm}
\begin{align}\nonumber
& \mathbf{A_1}=\left(\begin{array}{l l} \lambda_{kl} \mathbf{I} \quad \mathbf{0} \\ \mathbf{0} \quad -\sigma_k^2-\lambda_{kl}\epsilon^2 \end{array}\right), \quad\mathbf{A_2} = \left(\begin{array}{l l} \mathbf{\nu}_k \quad \mathbf{0} \\ \mathbf{0} \quad\;\, 0 \end{array}\right).
\end{align}

\vspace{-0.2cm}
The following KKT conditions hold for \eqref{Procedure2} 

\vspace{-0.5cm}
\begin{align}\nonumber
&\frac{\partial \mathcal{L}}{\partial \mathbf{W}_k}=\mathbf{0} \Rightarrow \mathbf{Y}_k+\mathbf{H}_{l}
 \mathbf{T}_{kl} \mathbf{H}^H_{l}/\gamma^{min}_k= \mathbf{I}\\ \label{kkt}
&\qquad\qquad\qquad+ \sum_{j=1}^{k-1}\mathbf{H}_{l}\mathbf{T}_{jl} \mathbf{H}^H_{l}
+\sum_{j=k+1}^{K} \mathbf{T}_{jl},\\
&\mathbf{W}_k\mathbf{Y}_k=\mathbf{0},\\ \label{kkt2}
&(\mathbf{A_1}+ \mathbf{H}^H_{l} \mathbf{\phi}_k \mathbf{H}_{l}+ \mathbf{A_{2}})\mathbf{T}_{kl}=\mathbf{0}.
\end{align}

We premultiply \eqref{kkt} by $\mathbf{W}_k$, i.e.,

\vspace{-0.5cm}
\begin{align}\label{kkt3}
&\mathbf{W}_k \mathbf{H}_{l}
 \mathbf{T}_{kl} \mathbf{H}^H_{l}/\gamma^{min}_k=\mathbf{W}_k \big(\mathbf{I} + \sum_{j=1}^{k-1}\mathbf{H}_{l}\mathbf{T}_{jl} \mathbf{H}^H_{l}
+ \sum_{j=k+1}^{K}\mathbf{T}_{jl} \big),
\end{align}

\vspace{-0.5cm}
\indent Then, we can write the following rank relation

\vspace{-0.6cm}
\begin{align}\nonumber
 \text{rank}(\mathbf{W}_k)&=\text{rank}\Big[\mathbf{W}_k \big(\mathbf{I}+\sum_{j=1}^{k-1}\mathbf{H}_{l}\mathbf{T}_{jl} \mathbf{H}^H_{l}
+\hspace{-0.2cm}\sum_{j=k+1}^{K} \mathbf{T}_{jl} \big)\Big]\\ \nonumber
 &=\text{rank}(\mathbf{W}_k \mathbf{H}_{l}
 \mathbf{T}_{kl} \mathbf{H}^H_{l})\\ \label{kkt4}
 &\leq \min \{ \text{rank}(\mathbf{H}_{l}
 \mathbf{T}_{kl} \mathbf{H}^H_{l}), \text{rank}(\mathbf{W}_k)\}.
\end{align}
\indent  Based on \eqref{kkt4}, it is required to show
$\text{rank}(\mathbf{H}_{l}
 \mathbf{T}_{kl} \mathbf{H}^H_{l})\leq1$ if we claim $\text{rank}(\mathbf{W}_k)\leq1$.

First, we consider the following equations and Lemma 3:

\begin{align}\nonumber
&  [\mathbf{I}\quad \mathbf{0}]\mathbf{H}^H_{l}=\mathbf{I}, ~~[\mathbf{I}\quad \mathbf{0}]\mathbf{A_1}=\lambda_{kl}(\mathbf{H}^H_{l}-[\mathbf{0}_M \quad \mathbf{h}_l]),\\ \label{equal3}
&[\mathbf{I}\quad \mathbf{0}]\mathbf{A_2}=\mathbf{\nu}_k(\mathbf{H}^H_{l}-[\mathbf{0}_M \quad \mathbf{h}_l]),
\end{align}

\vspace{-0.2cm}
\textit{Lemma 3:} If a block Hermitian matrix \mbox{$\mathbf{B}=\hspace{-0.1cm}\left[\hspace{-0.2cm}\begin{array}{l l} \mathbf{B_1} \quad \mathbf{B_2}\\ \mathbf{B_3}\quad \mathbf{B_4}\end{array}\hspace{-0.2cm}\right]\hspace{-0.1cm}\succeq 0$}
then the main diagonal matrices $\mathbf{B_1} $ and $\mathbf{B_4} $ must be positive
definite (PSD) matrices \cite{sproceturebook}. 

%

We pre-multiply $[\mathbf{I}\quad \mathbf{0}]$ and post-multiply $\mathbf{H}^H_{l}$
by \eqref{kkt2}, respectively, and applying the equalities in \eqref{equal3}:

\vspace{-0.5cm}
\begin{align}\nonumber
& \lambda_{kl}(\mathbf{H}^H_{l}-[\mathbf{0}_M ~ \mathbf{h}_l])\mathbf{T}_{kl} \mathbf{H}^H_{l}+\mathbf{\nu}_k(\mathbf{H}^H_{l}-[\mathbf{0}_M ~ \mathbf{h}_l])\mathbf{T}_{kl} \mathbf{H}^H_{l}  \\\nonumber
&+\mathbf{\phi}_k\mathbf{H}_{l} \mathbf{T}_{kl} \mathbf{H}^H_{l}=\mathbf{0}\Rightarrow\\\label{prof}
&\!\!(\lambda_{kl}\mathbf{I}\!+\!\mathbf{\phi}_k\!+\!\mathbf{\nu}_k)\mathbf{H}_{l}\mathbf{T}_{kl}\mathbf{H}^H_{l}\!\!\!\!=\!\!(\lambda_{kl}\mathbf{I}+\mathbf{\nu}_k)[\mathbf{0}_M ~ \!\mathbf{h}_l]\mathbf{T}_{kl} \mathbf{H}^H_{l}
\end{align}

By applying Lemma 3 to \eqref{const1}, we can claim $(\lambda_{kl}\mathbf{I}+\mathbf{\phi}_k+\mathbf{\nu}_k)\succeq \mathbf{0}$
and is nonsingular; thus, multiplying by a nonsingular
matrix will not change the matrix rank. Thus, the following
rank relation holds:

\vspace{-0.6cm}
\begin{align} \nonumber
\text{rank}(\mathbf{H}_{l}
 \mathbf{T}_{kl} \mathbf{H}^H_{l})=&\text{rank}((\lambda_{kl}\mathbf{I}+\mathbf{\nu}_k)[\mathbf{0} \quad \mathbf{h}_l]
 \mathbf{T}_{kl} \mathbf{H}^H_{l})\\
& \leq \text{rank}([\mathbf{0} \quad \mathbf{h}_l])\leq 1.
\end{align}

\vspace{-0.1cm}
This completes the proof of Lemma 2. \qquad\qquad\qquad $\blacksquare$


%

\vspace{-0.3cm}
\bibliographystyle{IEEEtrans}
\bibliography{IEEEabrv,Bibliography}
\end{document}